\documentclass[prl,showpacs,amsfonts,amsmath,twocolumn,superscriptaddress]{revtex4-1}
\usepackage{array,amsmath,amsfonts,amssymb,dsfont}
\usepackage{natbib}
\usepackage[T1]{fontenc}
\usepackage{ae}
\usepackage{color}
\usepackage{bbold}
\usepackage{graphicx}
\usepackage[caption=false]{subfig}

\def\1{\mathchoice{\rm 1\mskip-4.2mu l}{\rm 1\mskip-4.2mu l}{\rm 1\mskip-4.6mu l}{\rm 1\mskip-5.2mu l}}

\newcommand{\mhz}{\ensuremath{\, \mathrm{MHz}}}
\newcommand{\ghz}{\ensuremath{\, \mathrm{GHz}}}

\newcommand{\ns}{\ensuremath{\, \mathrm{ns}}}
\newcommand{\mk}{\ensuremath{\, \mathrm{mK}}}
\newcommand{\K}{\ensuremath{\, \mathrm{K}}}
\newcommand{\um}{\ensuremath{\, \mathrm{\mu m}}}
\newcommand{\mm}{\ensuremath{\, \mathrm{mm}}}

\newcommand{\mT}{\ensuremath{\, \mathrm{mT}}}

\begin{document}

\title{Time-resolved measurements of surface spin-wave pulses at millikelvin temperatures}
\author{A.\ F.\ van Loo}
 \affiliation{Clarendon Laboratory, Department of Physics, University of Oxford, OX1 3PU, Oxford, United Kingdom}
\author{R.\ G.\ E.\ Morris}
 \affiliation{Clarendon Laboratory, Department of Physics, University of Oxford, OX1 3PU, Oxford, United Kingdom}
\author{A.\ D.\ Karenowska}
 \affiliation{Clarendon Laboratory, Department of Physics, University of Oxford, OX1 3PU, Oxford, United Kingdom}
 \email{alexy.karenowska@physics.ox.ac.uk}
\date{\today}

\pacs{}

\begin{abstract} 
    We experimentally investigate the propagation of pulsed magnetostatic surface spin-wave (magnon) signals in an yttrium iron garnet (YIG) waveguide at millikelvin temperatures. Our measurements are performed in a dilution refrigerator at microwave frequencies. The excellent signal-to-noise ratio afforded by the low-temperature environment allows the propagation of the pulses to be observed in detail. The evolution of the envelope shape as the spin-wave travels is found to be consistent with calculations based on the known dispersion relation for YIG. We observe a temperature-dependent shift of the ferromagnetic resonance frequency below 4K which we suggest is due to the low-temperature properties of the substrate below the film, gallium gadolinium garnet. Our measurement and the accompanying calculations give insight into both low-temperature magnon dynamics in YIG and the feasibility of the use of propagating magnons in solid-state quantum information processing.
\end{abstract}
\maketitle

The unusual and highly tunable dispersion of propagating magnons in yttrium iron garnet ($\mathrm{Y}_3\mathrm{Fe}_5\mathrm{O}_{12}$; YIG) has recently excited the interest of the superconducting circuit quantum electrodynamics (QED) community. The frequencies at which superconducting quantum devices typically operate overlap with the band in which magnons can be excited in YIG and, as superconducting quantum bit (qubit) structures are generally sensitive to electromagnetic fields, magnons and superconducting qubits can be made to communicate. So far, magnons in spheres of YIG have been shown to couple strongly to three-dimensional cavities~\cite{Tabuchi2014, Bourhill2016, Zhang2014, Zhang2015}, reentrant cavities~\cite{Goryachev2014, Kostylev2016}, and superconducting qubits via such cavities~\cite{Tabuchi2015}. Furthermore, magnons in YIG films have been shown to interact with superconducting coplanar waveguides and three-dimensional resonators~\cite{Huebl2013, Zhang2016}.

In this work, we investigate the propagation of magnetostatic surface magnons in a YIG waveguide made from a high-purity monocrystalline film grown by liquid-phase epitaxy on a gallium gadolinium garnet (GGG) substrate. Our measurements are performed in a dilution refrigerator at approximately 20 mK. At this temperature, thermal excitations of gigahertz-frequency magnons and photons are negligible. We previously investigated a similar system \cite{Karenowska2015}, focusing on achieving signal limits equivalent to the propagation of a single magnon. In this paper, we study the propagation of the excitations in significantly greater detail, comparing our experimental measurements with the predictions of theory. Understanding low-temperature magnon propagation is an essential step toward the integration of systems of propagating magnons with circuit QED systems. Work in this area sets its sights not only on the development of new quantum devices but also on the use of the sophisticated technology and methods developed for superconducting quantum computing to investigate the physics of single magnons.

To excite propagating magnons in a waveguide, a magnetic bias field must be applied. The dynamics of the particular modes that can be observed depend on the orientation of the bias field relative to the waveguide axis: in this study, we focus on magnetostatic surface spin waves (MSSWs; also called ``Damon-Eshbach modes'') that propagate perpendicular to an in-plane magnetic field.

The dispersion relation for MSSWs \cite{Kalinikos1981} can be written as

\begin{equation}
    \omega = \sqrt{\omega_H \left( \omega_H + \omega_M \right) + \frac{\omega_M^2}{4} \left( 1- \mathrm{e}^{-2 k d} \right)}
    \label{eq:dispersion}
\end{equation}

\noindent where $\omega_M = - \gamma \mu_0 M_\text{S}$ and $\omega_H = -\gamma B$. Here, $d$ is the film thickness, $\gamma$ is the gyromagnetic ratio, $\mu_0$ is the vacuum permeability, $B$ is the applied magnetic field, and $M_\text{S}$ the saturation magnetization ($197.4 \,\text{kA\,m}^{-1}$ in our film at approximately 20 mK). From the dispersion relation, the group and phase velocities ($v_\text{gr}$ and $v_\text{ph}$ respectively) can be derived:

\begin{equation}
    v_\text{gr} = \frac{\partial \omega}{ \partial k} = \frac{d \left( ( 2 \omega_H + \omega_M )^2 - 4 \omega^2 \right)}{4 \sqrt{\omega^2}}
    \label{eq:groupVelocity}
\end{equation}
\begin{equation}
    v_\text{ph} = \frac{\omega }{ k}=\frac{-2d \sqrt{\omega^2}}{\log{\left[\frac{(-2 \omega + 2 \omega_H + \omega_M)(2(\omega + \omega_H) + \omega_M)}{\omega_M^2}\right]}}
    \label{eq:phaseVelocity}
\end{equation}

\begin{figure}[!t]
    \centering
    \includegraphics[width=8.6cm]{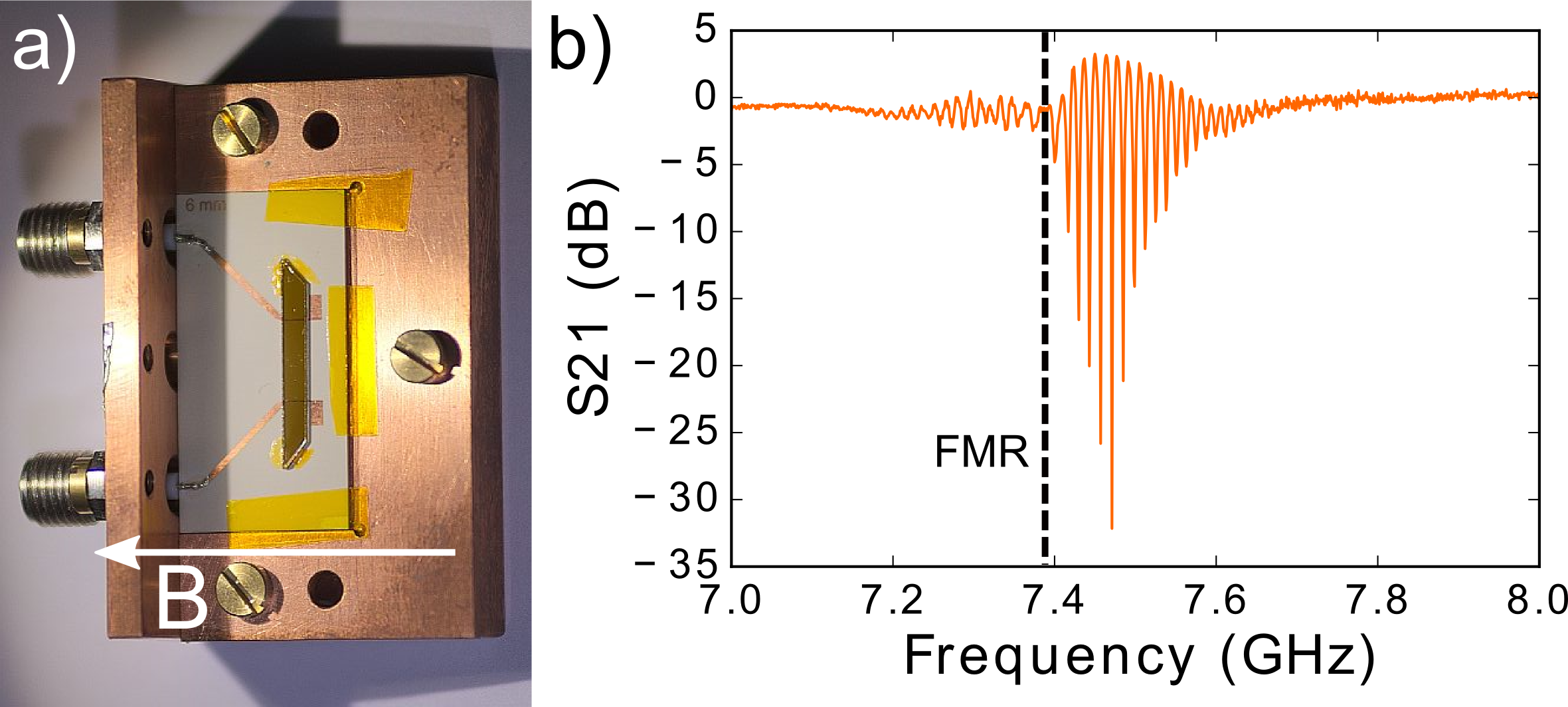}
    \caption{(a) The YIG waveguide is attached to a printed circuit board with two inductive antennae on its surface (an input and an output). The complete experimental assembly is housed in a copper box attached to the cold plate of a dilution refrigerator. (b) Measured transmission of a continuous-wave signal through the sample in a magnetic bias field of $167\, \text{mT}$. The dashed line indicates the position of the ferromagnetic resonance frequency.}
    \label{fig:fig1}
\end{figure}

The YIG waveguide used in our experiments is 9\um~thick, $2\,\text{mm}$ wide, and approximately $15\,\text{mm}$ long. To prevent coherent reflections of spin waves from the ends, both are cleaved at a 45 degree angle. The waveguide is affixed to a PCB in close contact with two lithographed stripline antennae [Fig.~\ref{fig:fig1}(a)] which are 50\um~wide and $6\,\text{mm}$ apart. The complete assembly is housed in a copper sample box attached to the cold plate of a dilution refrigerator. A superconducting magnet is connected to the $4\text{K}$ plate of the refrigerator and positioned so that the homogeneity of the field it produces is maximal at the position of the waveguide. The orientation of the field relative to the waveguide [Fig.~\ref{fig:fig1}(a)] is such that only MSSWs are excited. Because of the nonreciprocal nature of these excitations, they are only able to travel in one direction on the film surface adjacent to the antennae: from the input port to the output port of the experiment \cite{Kalinikos1981, Gurevich1996}.

\begin{figure}[!t]
    \centering
    \includegraphics[width=8.6cm]{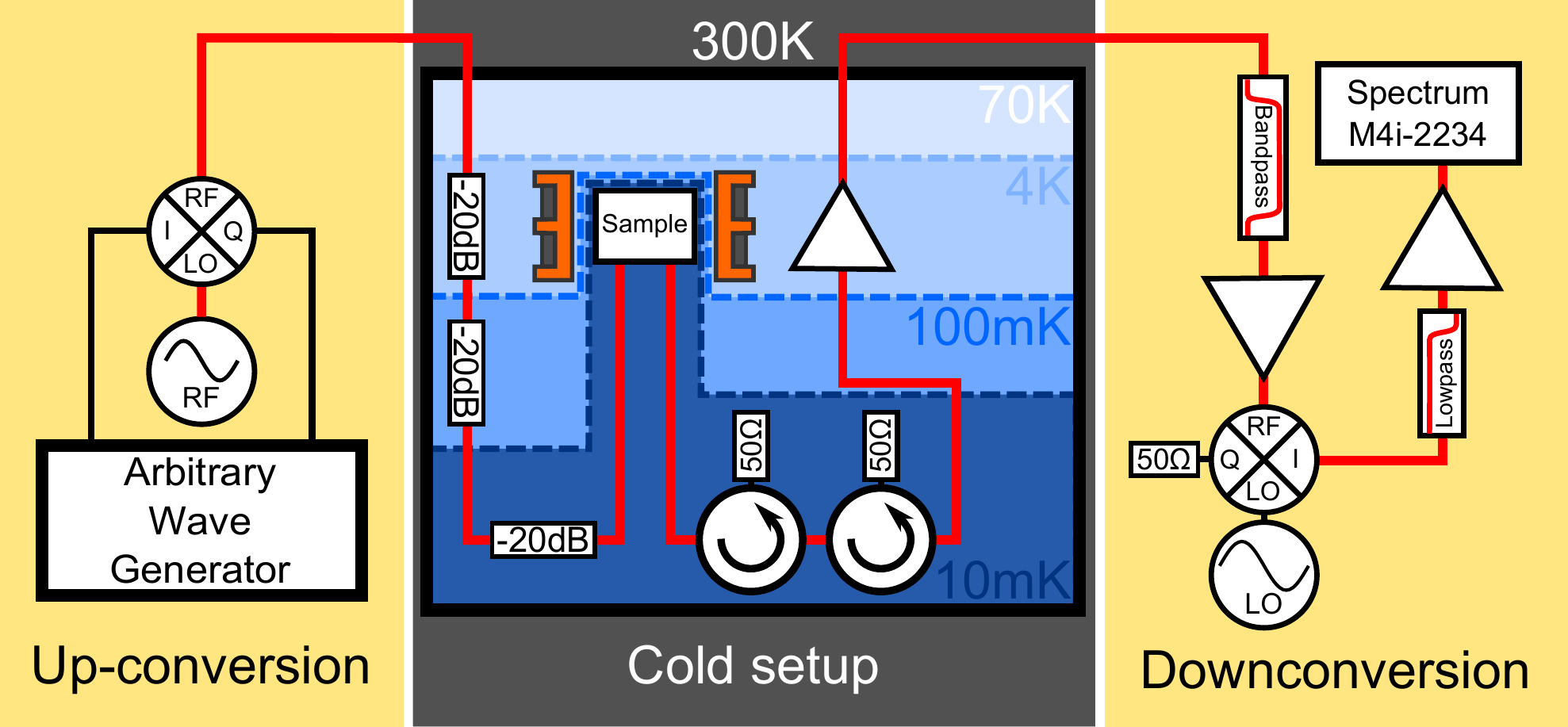}
    \caption{The microwave setup used in our experiments can be divided into three parts. In the up-conversion section, short microwave pulses are created by mixing the output of a continuous-wave microwave source with envelope shapes generated by an arbitrary-waveform generator. Inside the dilution refrigerator, the input line to the sample is heavily attenuated to thermalize the signal, such that the electronic noise temperature is decreased to a level similar to the phonon noise temperature. The down-conversion system allows the recovery of the envelope of the signal transmitted through the waveguide.}
    \label{fig:fig2}
\end{figure}

To ensure that room-temperature noise cannot couple into the waveguide, the line connected to the input antenna is heavily attenuated inside the refrigerator [Fig.~\ref{fig:fig2}]. The output antenna is connected to a low-temperature HEMT amplifier via two isolators. The amplified signal is filtered and down-converted to an intermediate frequency of $500\,\mhz$ at room temperature. A fast data acquisition card (Spectrum M4i-2234-x8) records the resulting voltage as a function of time at a rate of $2.5\,\ghz$ for $800\,\ns$. Typically $2^{11}$ to $2^{16}$ such traces are recorded and averaged, although much higher numbers are can be used if desired (e.g.\ when the input signal is reduced to very low levels). Through this step, incoherent and inelastically scattered signals cancel out, leaving only the coherently scattered radiation at the drive frequency. After a digital down-conversion and filtering step, the envelope of the transmitted microwave signal is recovered.

When a microwave-frequency signal is applied to the input antenna, there are two routes by which it can reach the output: via the vacuum in the sample box as microwave photons at the speed of light $c$, and as MSSWs in the magnetic waveguide at the group velocity determined by eq. \ref{eq:groupVelocity}. The MSSW group velocity is typically several orders of magnitude slower than the speed of light.

To perform an initial characterization of the YIG film, we connect the cold setup [center part in Fig.~\ref{fig:fig2}] to a network analyzer and measure the transmission (S21) of the experimental system between $4$ and $8\, \text{GHz}$. Results from such a measurement are shown in Fig.~\ref{fig:fig1}(b). The low-frequency limit of the magnon passband is the ferromagnetic resonance (FMR) frequency (around $7.4\, \text{GHz}$ in Fig.~\ref{fig:fig1}(b)); this is the $k=0$ mode of the magnon system in which the spins throughout the YIG film precess in-phase. The oscillations in the signal transmittance as a function of frequency are caused by interference between the vacuum and magnon signals.

Having performed our initial characterization in the continuous-wave regime, we investigate the dynamics of the magnetic system by driving it with short microwave pulses. The difference in propagation speed between the MSSW and the signal propagating through the vacuum here works to our advantage, allowing us to separate these two responses in time for sufficiently short pulses. The short microwave pulses also have a relatively wide spectral bandwidth, and as such allow the investigation of a range of magnetostatic surface spin-waves with different wave numbers at a single magnetic field. Figure \ref{fig:fig2} shows the experimental setup used to generate and measure the transmittance of short microwave pulses through the YIG sample.

\begin{figure}[!t]
    \centering
    \includegraphics[width=8.6cm]{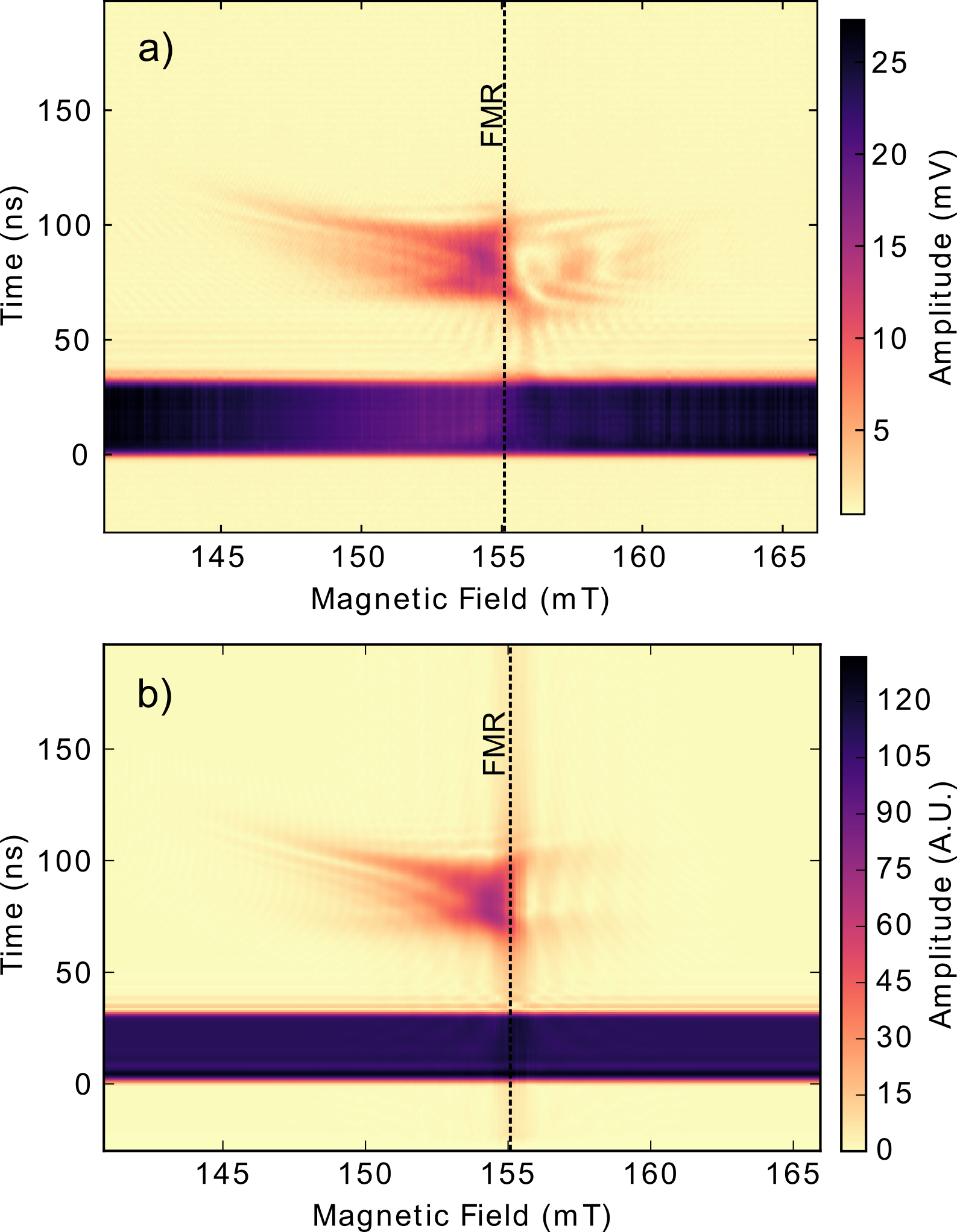}
    \caption{Measurements and simulations of pulsed propagating magnon signals. (a) Experimental data showing the transmitted signal amplitude due to a square $30\,\text{ns}$ microwave pulse with a carrier frequency of $7\, \text{GHz}$. The magnetic bias field (horizontal axis) is swept across the magnon band. (b) Results of a simulation under the same conditions.}
    \label{fig:fig3}
\end{figure}

Figure \ref{fig:fig3}a shows the spin-wave output signal in response to the application of a square $30\,\text{ns}$ microwave pulse of constant carrier frequency ($7\, \text{GHz}$) as the magnetic bias field is swept across the magnon band. In oder to investigate magnons with a variety of k-values, we can either keep the bias field constant and vary the carrier frequency, or keep the carrier frequency constant and sweep the bias field. Here we choose to sweep the magnet current at a fixed carrier frequency to eliminate the effects that the frequency-dependent attenuation of the microwave lines and components have on the data. The horizontal stripe of high signal intensity from $0$ to $30\,\text{ns}$ is due to direct electromagnetic interaction between the two antennae. The magnon signal begins to appear at approximately $70\,\text{ns}$. For each vertical slice, the difference in shape between the initial pulse and the magnon response is due to the different frequency components within the pulse traveling at different velocities, given by the dispersion relation. For each slice, therefore, we sample the dispersion relation with a simultaneous range of frequencies given by the spectrum of our input pulse, which is limited by the $300\mhz$ output bandwidth of our arbitrary-waveform generator. The field at which FMR occurs is $\text{155}\, \text{mT}$. For clarity, the FMR position is indicated in Fig.~\ref{fig:fig3}. When the field is reduced below that corresponding to the FMR frequency (moving from right to left along the horizontal axis in Fig.~\ref{fig:fig3}), the carrier frequency in the pulse excites modes of increasingly higher $k$ which have a lower group velocity [Eq.~\ref{eq:groupVelocity}] and therefore take longer to reach the output antenna. As the field is reduced further, the region of the dispersion relation sampled by the input pulse is shifted. The power used for these measurements is on the order of $10^{5}$ photons per pulse at the input antenna. We previously studied the possibility of reducing the power to the single-magnon power level in our setup for both thin films \cite{Karenowska2015} and YIG spheres\cite{Morris2017}.

Figure~\ref{fig:fig3}(b) shows the results of a simulation of the response of the experimental system calculated on the basis of Eq.~\ref{eq:dispersion}. These calculations are performed as follows: Like the experimental pulse, the simulated input signal is constructed by our multiplying a continuous-wave carrier signal with a $30\,\ns$ square envelope. A very small offset is added to the square pulse (before multiplication) to account for imperfect mixer calibration in the experiment. This input signal is Fourier transformed and, for each frequency component, the phase velocity is calculated. The arrival time of each frequency component at the output antenna is then determined from the phase velocity and the distance between the antennae. A delay in time is equivalent to a phase shift in Fourier space, and thus the $k$-dependent travel time is implemented by multiplying the Fourier component by $e^{-i\omega d/ v_p}$, $d$ being the inter-antenna distance ($6\, \mm$ for the sample in Fig.~\ref{fig:fig3}), and $v_p$ the phase velocity of the magnon at that particular frequency. The propagation loss of surface waves is expected to be a function of time only \cite{Stancil2009}. Therefore, waves of different $k$ values experience different amounts of loss when traveling the same inter-antenna distance because of their different phase velocities. This is taken into account in the simulations by multiplying the Fourier components by $e^{-d/(v_p T_{L})}$, where $T_{L}$ is the characteristic frequency-independent \cite{Stancil2009} loss time. The time-dependent signal at the output antenna is reconstructed via an inverse Fourier transform, after which the analog and digital signal processing applied to the data as described above is mimicked. To avoid overfitting, the antennae and sample box are treated as being ideal, save for a $k$-dependent coupling factor between the antennae and the YIG, which we take, according to \cite{Stancil2009}, to be of the form $J_0(k w/2)$, where $J_0$ is the zeroth-order Bessel function. The ratio between antenna-YIG coupling and antenna-antenna coupling is tuned manually to resemble the data.

As predicted by eq.~\ref{eq:groupVelocity}, the speed of the magnons reduces as their wavenumber increases, resulting in an upward curve of the magnon signal on the left side of Figs.~\ref{fig:fig3}(a) and \ref{fig:fig3}(b). For these data, the bandwidth of the magnon signal (measured by the range of magnetic fields for which we observe a magnon signal at a constant frequency) is not limited by the bandwidth of the antenna. Rather, the width of the measurable magnon band is limited by the propagation loss rate of magnons traveling in YIG. This was confirmed by measuring a similar system with a smaller inter-antenna distance that is found to have a higher signal bandwidth. Fitting our simulations to the data, we find that the characteristic loss time $T_L=0.85\, \pm .05\, \ns$. Note that this does not mean that magnons typically decay within this time: $T_L$ is referenced to the phase rather than the group velocity and is used to model the different amounts of loss that -- by virtue of the different phase velocities -- the different frequency components are subject to. The typical timescale for the pulse to lose $1/e$ of its energy is approximately $30\,\ns$. Overall, it can be seen that there is good agreement between experiment and theory, especially regarding the structure and shape of the magnon response to the left of the FMR in the plot. There are, however, also some differences that warrant discussion. 

One difference between simulation and experiment is the vertical line close to the FMR frequency seen only in the theoretical plot. This line is an artifact of the simulation that comes about from the phase velocity tending to infinity as $k$ reduces to zero. In the real physical system the phase velocity diverges close to the FMR frequency, producing a signal between the magnon and the directly-coupled pulses. We also see some ringing in the experiments, appearing as low-amplitude oscillations following the directly-coupled signal. This feature is not exactly reproduced in the simulations and occurs as a result of unmodeled nonidealities of the antennae and sample box. A further difference exists at magnetic fields above $155\,\mT$: in Fig.~\ref{fig:fig3}(a) there is a structure in the measurement that is not reproduced in the simulations. This signal disappears when the magnetic bias field is reversed, confirming it must be due to a surface mode rather than a (reciprocal) backward or forward volume excitation. We have investigated several effects as potential origins of this signal, including the finite width of the waveguide and the inhomogeneity of the magnetic bias field. To exclude the effect of a finite waveguide width as an explanation, we used a modified dispersion relation given in Ref.~\cite{Okeeffe1978} to calculate the phase velocity in our simulations. An inhomogeneous magnetic bias field was implemented in our model by performing the simulations for a small range of fields simultaneously, mimicking the effects of different parts of the YIG film being subject to different bias fields. Neither of these alterations to the simulations produced the signal observed in Fig.~\ref{fig:fig3} at fields above 155~\mT. Effects due to the presence of a ground plane close to the sample, which can be calculated from Ref.~\cite{yukawa1977}, are expected to be very small for our experimental setup, and likewise did not produce the observed signal when they were included in the simulations. A possible explanation for this sub-FMR signal is that the inhomogeneous magnetic fields close to the edges of the waveguide cause localized modes \cite{Guslienko2003}. It should be noted that evidence of this signal is also found in the continuous-wave experiment [to the left of the FMR line in Fig.~\ref{fig:fig1}b], and similar signals have also been reported in room-temperature data in the literature (e.g., Ref.~\cite{Serga2010}).

Data such as shown in Fig.~\ref{fig:fig3}(a) were obtained at several carrier frequencies and powers across the measurement bandwidth of our system (between 4 and 8\ghz). Figure~\ref{fig:fig3} is representative in both the agreements and disagreements between the experiments and simulations. 

\begin{figure}[!h]
    \centering
    \includegraphics[width=8.6cm]{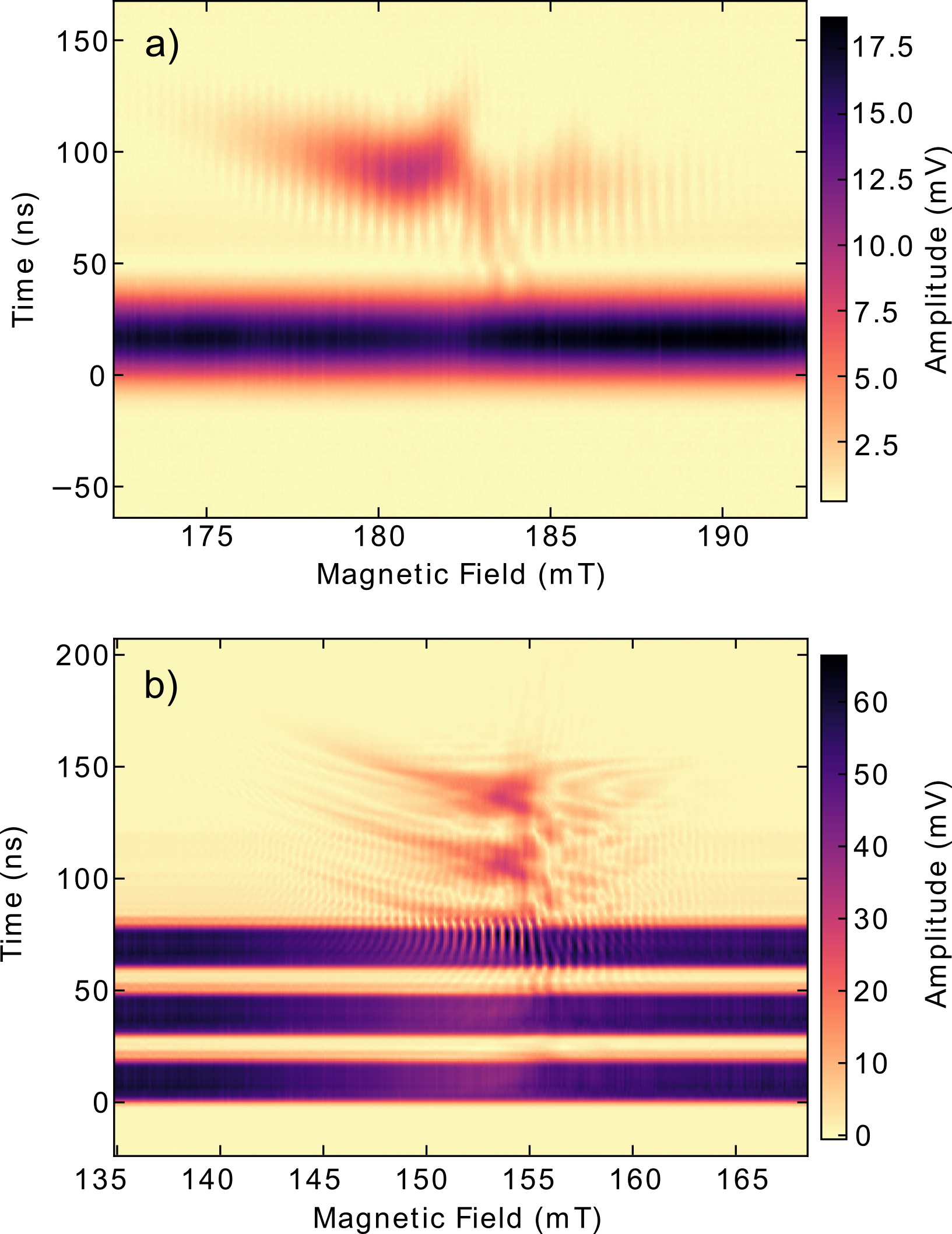}
    \caption{Experimental data showing the propagation of (a) a Gaussian pulse with $\sigma=12\,\ns$ and (b) a train of three $20\,\ns$~pulses separated by $10\,\ns$.}
    \label{fig:fig4}
\end{figure}

Our microwave setup can be used to inject pulses of arbitrary envelope into the experimental system. In Fig.~\ref{fig:fig4}(a), we see the output signal in response to a Gaussian input with $\sigma=12\,\ns$, and in Fig.~\ref{fig:fig4}(b), a train of three $20\,\ns$ pulses separated by $10\,\ns$. In Fig.~\ref{fig:fig4}(b), the signal produced by the direct coupling of the last of the input pulses from the input antenna to the output antenna overlaps with the first magnon pulse. The result is an interference pattern akin to that seen in Fig.~\ref{fig:fig1}(b). The clear separation between the individual pulses in the train underlines the promising potential of magnon systems as a platform for the transmission of information signals. 

\begin{figure}[!h]
    \centering
    \includegraphics[width=8.6cm]{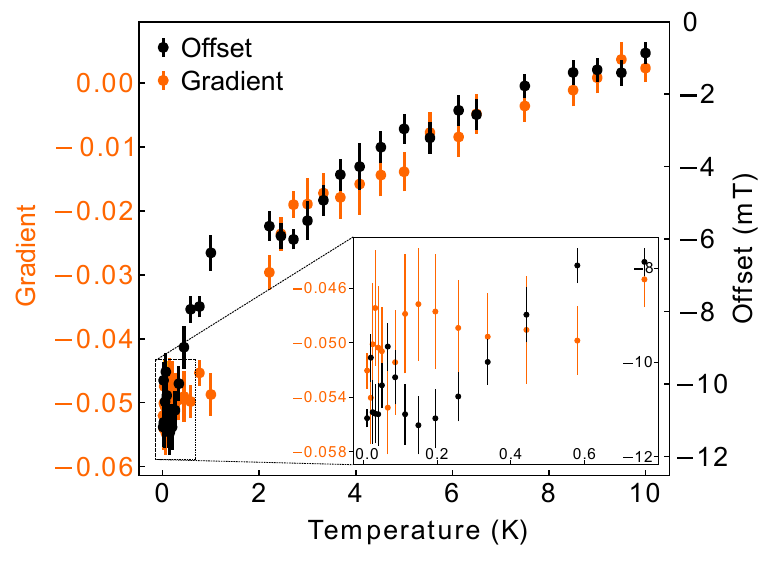}
    \caption{At each temperature, the deviation of the observed FMR frequency from the theoretical value can be fitted as a linear function of the applied magnetic field. The offset and gradient of these linear functions are plotted here for a range of different temperatures.}
    \label{fig:fig5}
\end{figure}

When comparing the FMR frequency with the frequency predicted by Eq.~\ref{eq:dispersion}, we find a slight difference between observation and theory. Repeating measurements similar to those shown in Fig.~\ref{fig:fig1}(b) for different magnetic fields, we find that the frequency difference as a function of the magnetic field is well fitted with a linear function. This shift in frequency could be due to the field at the YIG being slightly different from the field that we apply, or the YIG FMR frequency being different from that predicted by Eq.~\ref{eq:dispersion}. If we assume that the field at the YIG is different from the applied field, we can use the difference between the expected and observed FMR to extract the offset magnetic field needed to explain this frequency difference. The total magnetic field is then given by  $B_{tot} = B_{mag} + \alpha B_{mag} + \beta$, with $\alpha=-0.052\pm 0.002$ and $\beta = -0.0112 \pm 0.0002 \mathrm{T}$. These values for $\alpha$ and $\beta$ are extracted from data taken at base temperature (approximately $20\,\mathrm{mK}$). The same procedure is used over a range of temperatures up to $10\,\mathrm{K}$ in our dilution refrigerator. The results are shown in Fig.~\ref{fig:fig5}. As can be seen in the inset, the values for $\alpha$ and $\beta$ do not measurably vary up to $0.5\,\mathrm{K}$. Beyond that, the magnitude of both fitted parameters reduces, seemingly converging toward zero at higher temperatures. The value of the YIG's saturation magnetization is expected to not change significantly at these temperatures \cite{Anderson1964, Hansen1974a}, and is here taken to be constant at $197.4\,\mathrm{kA/m}$, . The values in the horizontal axes of Fig.~\ref{fig:fig3}(a) and Fig.~\ref{fig:fig4} were adjusted such that $B_{tot}$ is shown rather than $B_{mag}$.

Our prime suspect for the measured deviation of the FMR frequency is the substrate for the YIG film, GGG. While GGG is a convenient magnetically inert material to use for spin-wave experiments at room temperature, at cryogenic temperatures down to $1.5\,\K$, it is paramagnetic, and displays Curie-Weiss susceptibility \cite{Schiffer1995}. In this regime the short relaxation time of the paramagnetic GGG dampens spin-wave propagation of YIG deposited on its surface. The exchange interaction between neighboring spins in paramagnetic GGG is approximately $1.5\,\K/k_{B}$ \cite{Schiffer1994}, and its behavior is therefore expected to be different below this temperature but is not well documented. There are suggestions that GGG undergoes a spin glass transition  around $200\,\mk$ \cite{Deen2015} or enters a complex magnetic state involving locally correlated spin loops \cite{Paddison2015}. While it is beyond the scope of this work to establish the low-temperature behaviour of GGG, assuming the observed shift in FMR to be caused by this material, we can make a few observations. First, $\alpha$ being nonzero should not surprise us as GGG is known to become paramagnetic at lower temperatures. The nonzero value of $\beta$, however, is more interesting: this would imply a magnetic ordering of the GGG substrate. Note that our measurements were not extended to zero field as we cannot measure the FMR frequency outside $4-8\, \mathrm{GHz}$, the band of our measurement setup.

We suggest that the relatively high measured loss experienced by the magnon signal during propagation is related to the observed low-temperature magnetic properties of GGG. Coupling YIG to a magnetic material with higher spin-wave damping certainly would be expected to result in such an effect. It should be noted that we ignore anisotropy in these considerations. Effects due to anisotropy are generally expected to be too small to account for the irregularities we record in our measurements, but as far as we are aware, this has not been experimentally confirmed at these temperatures.

In conclusion, in this work we present a detailed study of the propagation of pulsed magnetostatic surface spin-wave signals in an yttrium iron garnet waveguide at millikelvin temperatures. Calculations using a simple model based on the dispersion relation of MSSWs agree well with our experimental findings. Our investigations confirm that trains of short spin-wave pulses can readily be excited and detected inductively in YIG waveguides grown on gallium gadolinium garnet substrates. We observe a temperature- and field-dependent deviation from the theoretically expected FMR frequency, which we attribute to the low-temperature magnetic ordering in the substrate GGG. As well as being of interest in their own right, these results are an important step toward the combination of systems of propagating magnons with microwave superconducting quantum circuit technology.

The authors acknowledge Bob Watkins for preparing the waveguide and the EPSRC for funding this research with Grant EP/K032690/1.

\bibliographystyle{apsrev4-1}
\bibliography{./paper1}

\begin{thebibliography}{24}%
\makeatletter
\providecommand \@ifxundefined [1]{%
 \@ifx{#1\undefined}
}%
\providecommand \@ifnum [1]{%
 \ifnum #1\expandafter \@firstoftwo
 \else \expandafter \@secondoftwo
 \fi
}%
\providecommand \@ifx [1]{%
 \ifx #1\expandafter \@firstoftwo
 \else \expandafter \@secondoftwo
 \fi
}%
\providecommand \natexlab [1]{#1}%
\providecommand \enquote  [1]{``#1''}%
\providecommand \bibnamefont  [1]{#1}%
\providecommand \bibfnamefont [1]{#1}%
\providecommand \citenamefont [1]{#1}%
\providecommand \href@noop [0]{\@secondoftwo}%
\providecommand \href [0]{\begingroup \@sanitize@url \@href}%
\providecommand \@href[1]{\@@startlink{#1}\@@href}%
\providecommand \@@href[1]{\endgroup#1\@@endlink}%
\providecommand \@sanitize@url [0]{\catcode `\\12\catcode `\$12\catcode
  `\&12\catcode `\#12\catcode `\^12\catcode `\_12\catcode `\%12\relax}%
\providecommand \@@startlink[1]{}%
\providecommand \@@endlink[0]{}%
\providecommand \url  [0]{\begingroup\@sanitize@url \@url }%
\providecommand \@url [1]{\endgroup\@href {#1}{\urlprefix }}%
\providecommand \urlprefix  [0]{URL }%
\providecommand \Eprint [0]{\href }%
\providecommand \doibase [0]{http://dx.doi.org/}%
\providecommand \selectlanguage [0]{\@gobble}%
\providecommand \bibinfo  [0]{\@secondoftwo}%
\providecommand \bibfield  [0]{\@secondoftwo}%
\providecommand \translation [1]{[#1]}%
\providecommand \BibitemOpen [0]{}%
\providecommand \bibitemStop [0]{}%
\providecommand \bibitemNoStop [0]{.\EOS\space}%
\providecommand \EOS [0]{\spacefactor3000\relax}%
\providecommand \BibitemShut  [1]{\csname bibitem#1\endcsname}%
\let\auto@bib@innerbib\@empty
\bibitem [{\citenamefont {Tabuchi}\ \emph {et~al.}(2014)\citenamefont
  {Tabuchi}, \citenamefont {Ishino}, \citenamefont {Ishikawa}, \citenamefont
  {Yamazaki}, \citenamefont {Usami},\ and\ \citenamefont
  {Nakamura}}]{Tabuchi2014}%
  \BibitemOpen
  \bibfield  {author} {\bibinfo {author} {\bibfnamefont {Y.}~\bibnamefont
  {Tabuchi}}, \bibinfo {author} {\bibfnamefont {S.}~\bibnamefont {Ishino}},
  \bibinfo {author} {\bibfnamefont {T.}~\bibnamefont {Ishikawa}}, \bibinfo
  {author} {\bibfnamefont {R.}~\bibnamefont {Yamazaki}}, \bibinfo {author}
  {\bibfnamefont {K.}~\bibnamefont {Usami}}, \ and\ \bibinfo {author}
  {\bibfnamefont {Y.}~\bibnamefont {Nakamura}},\ }\href {\doibase
  10.1103/PhysRevLett.113.083603} {\bibfield  {journal} {\bibinfo  {journal}
  {Phys. Rev. Lett.}\ }\textbf {\bibinfo {volume} {113}},\ \bibinfo {pages}
  {083603} (\bibinfo {year} {2014})}\BibitemShut {NoStop}%
\bibitem [{\citenamefont {Bourhill}\ \emph {et~al.}(2016)\citenamefont
  {Bourhill}, \citenamefont {Kostylev}, \citenamefont {Goryachev},
  \citenamefont {Creedon},\ and\ \citenamefont {Tobar}}]{Bourhill2016}%
  \BibitemOpen
  \bibfield  {author} {\bibinfo {author} {\bibfnamefont {J.}~\bibnamefont
  {Bourhill}}, \bibinfo {author} {\bibfnamefont {N.}~\bibnamefont {Kostylev}},
  \bibinfo {author} {\bibfnamefont {M.}~\bibnamefont {Goryachev}}, \bibinfo
  {author} {\bibfnamefont {D.~L.}\ \bibnamefont {Creedon}}, \ and\ \bibinfo
  {author} {\bibfnamefont {M.~E.}\ \bibnamefont {Tobar}},\ }\href {\doibase
  10.1103/PhysRevB.93.144420} {\bibfield  {journal} {\bibinfo  {journal} {Phys.
  Rev. B}\ }\textbf {\bibinfo {volume} {93}},\ \bibinfo {pages} {144420}
  (\bibinfo {year} {2016})}\BibitemShut {NoStop}%
\bibitem [{\citenamefont {Zhang}\ \emph {et~al.}(2014)\citenamefont {Zhang},
  \citenamefont {Zou}, \citenamefont {Jiang},\ and\ \citenamefont
  {Tang}}]{Zhang2014}%
  \BibitemOpen
  \bibfield  {author} {\bibinfo {author} {\bibfnamefont {X.}~\bibnamefont
  {Zhang}}, \bibinfo {author} {\bibfnamefont {C.-L.}\ \bibnamefont {Zou}},
  \bibinfo {author} {\bibfnamefont {L.}~\bibnamefont {Jiang}}, \ and\ \bibinfo
  {author} {\bibfnamefont {H.~X.}\ \bibnamefont {Tang}},\ }\href {\doibase
  10.1103/PhysRevLett.113.156401} {\bibfield  {journal} {\bibinfo  {journal}
  {Phys. Rev. Lett.}\ }\textbf {\bibinfo {volume} {113}},\ \bibinfo {pages}
  {156401} (\bibinfo {year} {2014})}\BibitemShut {NoStop}%
\bibitem [{\citenamefont {Zhang}\ \emph {et~al.}(2015)\citenamefont {Zhang},
  \citenamefont {Wang}, \citenamefont {Li}, \citenamefont {Luo}, \citenamefont
  {Wu}, \citenamefont {Nori},\ and\ \citenamefont {You}}]{Zhang2015}%
  \BibitemOpen
  \bibfield  {author} {\bibinfo {author} {\bibfnamefont {D.}~\bibnamefont
  {Zhang}}, \bibinfo {author} {\bibfnamefont {X.-M.}\ \bibnamefont {Wang}},
  \bibinfo {author} {\bibfnamefont {T.-F.}\ \bibnamefont {Li}}, \bibinfo
  {author} {\bibfnamefont {X.-Q.}\ \bibnamefont {Luo}}, \bibinfo {author}
  {\bibfnamefont {W.}~\bibnamefont {Wu}}, \bibinfo {author} {\bibfnamefont
  {F.}~\bibnamefont {Nori}}, \ and\ \bibinfo {author} {\bibfnamefont {J.~Q.}\
  \bibnamefont {You}},\ }\href@noop {} {\bibfield  {journal} {\bibinfo
  {journal} {Npj Quantum Inf.}\ }\textbf {\bibinfo {volume} {1}},\ \bibinfo
  {pages} {15014} (\bibinfo {year} {2015})}\BibitemShut {NoStop}%
\bibitem [{\citenamefont {Goryachev}\ \emph {et~al.}(2014)\citenamefont
  {Goryachev}, \citenamefont {Farr}, \citenamefont {Creedon}, \citenamefont
  {Fan}, \citenamefont {Kostylev},\ and\ \citenamefont
  {Tobar}}]{Goryachev2014}%
  \BibitemOpen
  \bibfield  {author} {\bibinfo {author} {\bibfnamefont {M.}~\bibnamefont
  {Goryachev}}, \bibinfo {author} {\bibfnamefont {W.~G.}\ \bibnamefont {Farr}},
  \bibinfo {author} {\bibfnamefont {D.~L.}\ \bibnamefont {Creedon}}, \bibinfo
  {author} {\bibfnamefont {Y.}~\bibnamefont {Fan}}, \bibinfo {author}
  {\bibfnamefont {M.}~\bibnamefont {Kostylev}}, \ and\ \bibinfo {author}
  {\bibfnamefont {M.~E.}\ \bibnamefont {Tobar}},\ }\href {\doibase
  10.1103/PhysRevApplied.2.054002} {\bibfield  {journal} {\bibinfo  {journal}
  {Phys. Rev. Applied}\ }\textbf {\bibinfo {volume} {2}},\ \bibinfo {pages}
  {054002} (\bibinfo {year} {2014})}\BibitemShut {NoStop}%
\bibitem [{\citenamefont {Kostylev}\ \emph {et~al.}(2016)\citenamefont
  {Kostylev}, \citenamefont {Goryachev},\ and\ \citenamefont
  {Tobar}}]{Kostylev2016}%
  \BibitemOpen
  \bibfield  {author} {\bibinfo {author} {\bibfnamefont {N.}~\bibnamefont
  {Kostylev}}, \bibinfo {author} {\bibfnamefont {M.}~\bibnamefont {Goryachev}},
  \ and\ \bibinfo {author} {\bibfnamefont {M.~E.}\ \bibnamefont {Tobar}},\
  }\href {\doibase http://dx.doi.org/10.1063/1.4941730} {\bibfield  {journal}
  {\bibinfo  {journal} {Appl. Phys. Lett.}\ }\textbf {\bibinfo {volume}
  {108}},\ \bibinfo {eid} {062402} (\bibinfo {year} {2016}),\
  http://dx.doi.org/10.1063/1.4941730}\BibitemShut {NoStop}%
\bibitem [{\citenamefont {Tabuchi}\ \emph {et~al.}(2015)\citenamefont
  {Tabuchi}, \citenamefont {Ishino}, \citenamefont {Noguchi}, \citenamefont
  {Ishikawa}, \citenamefont {Yamazaki}, \citenamefont {Usami},\ and\
  \citenamefont {Nakamura}}]{Tabuchi2015}%
  \BibitemOpen
  \bibfield  {author} {\bibinfo {author} {\bibfnamefont {Y.}~\bibnamefont
  {Tabuchi}}, \bibinfo {author} {\bibfnamefont {S.}~\bibnamefont {Ishino}},
  \bibinfo {author} {\bibfnamefont {A.}~\bibnamefont {Noguchi}}, \bibinfo
  {author} {\bibfnamefont {T.}~\bibnamefont {Ishikawa}}, \bibinfo {author}
  {\bibfnamefont {R.}~\bibnamefont {Yamazaki}}, \bibinfo {author}
  {\bibfnamefont {K.}~\bibnamefont {Usami}}, \ and\ \bibinfo {author}
  {\bibfnamefont {Y.}~\bibnamefont {Nakamura}},\ }\href {\doibase
  10.1126/science.aaa3693} {\bibfield  {journal} {\bibinfo  {journal}
  {Science}\ } (\bibinfo {year} {2015}),\ 10.1126/science.aaa3693}\BibitemShut
  {NoStop}%
\bibitem [{\citenamefont {Huebl}\ \emph {et~al.}(2013)\citenamefont {Huebl},
  \citenamefont {Zollitsch}, \citenamefont {Lotze}, \citenamefont {Hocke},
  \citenamefont {Greifenstein}, \citenamefont {Marx}, \citenamefont {Gross},\
  and\ \citenamefont {Goennenwein}}]{Huebl2013}%
  \BibitemOpen
  \bibfield  {author} {\bibinfo {author} {\bibfnamefont {H.}~\bibnamefont
  {Huebl}}, \bibinfo {author} {\bibfnamefont {C.~W.}\ \bibnamefont
  {Zollitsch}}, \bibinfo {author} {\bibfnamefont {J.}~\bibnamefont {Lotze}},
  \bibinfo {author} {\bibfnamefont {F.}~\bibnamefont {Hocke}}, \bibinfo
  {author} {\bibfnamefont {M.}~\bibnamefont {Greifenstein}}, \bibinfo {author}
  {\bibfnamefont {A.}~\bibnamefont {Marx}}, \bibinfo {author} {\bibfnamefont
  {R.}~\bibnamefont {Gross}}, \ and\ \bibinfo {author} {\bibfnamefont
  {S.~T.~B.}\ \bibnamefont {Goennenwein}},\ }\href {\doibase
  10.1103/PhysRevLett.111.127003} {\bibfield  {journal} {\bibinfo  {journal}
  {Phys. Rev. Lett.}\ }\textbf {\bibinfo {volume} {111}},\ \bibinfo {pages}
  {127003} (\bibinfo {year} {2013})}\BibitemShut {NoStop}%
\bibitem [{\citenamefont {Zhang}\ \emph {et~al.}(2016)\citenamefont {Zhang},
  \citenamefont {Zou}, \citenamefont {Jiang},\ and\ \citenamefont
  {Tang}}]{Zhang2016}%
  \BibitemOpen
  \bibfield  {author} {\bibinfo {author} {\bibfnamefont {X.}~\bibnamefont
  {Zhang}}, \bibinfo {author} {\bibfnamefont {C.}~\bibnamefont {Zou}}, \bibinfo
  {author} {\bibfnamefont {L.}~\bibnamefont {Jiang}}, \ and\ \bibinfo {author}
  {\bibfnamefont {H.~X.}\ \bibnamefont {Tang}},\ }\href {\doibase
  http://dx.doi.org/10.1063/1.4939134} {\bibfield  {journal} {\bibinfo
  {journal} {J. Appl. Phys.}\ }\textbf {\bibinfo {volume} {119}},\ \bibinfo
  {eid} {023905} (\bibinfo {year} {2016}),\
  http://dx.doi.org/10.1063/1.4939134}\BibitemShut {NoStop}%
\bibitem [{\citenamefont {Karenowska}\ \emph {et~al.}(2015)\citenamefont
  {Karenowska}, \citenamefont {Patterson}, \citenamefont {Peterer},
  \citenamefont {Magn\'usson},\ and\ \citenamefont {Leek}}]{Karenowska2015}%
  \BibitemOpen
  \bibfield  {author} {\bibinfo {author} {\bibfnamefont {A.~D.}\ \bibnamefont
  {Karenowska}}, \bibinfo {author} {\bibfnamefont {A.~D.}\ \bibnamefont
  {Patterson}}, \bibinfo {author} {\bibfnamefont {M.~J.}\ \bibnamefont
  {Peterer}}, \bibinfo {author} {\bibfnamefont {E.~B.}\ \bibnamefont
  {Magn\'usson}}, \ and\ \bibinfo {author} {\bibfnamefont {P.~J.}\ \bibnamefont
  {Leek}},\ }\href@noop {} {\bibfield  {journal} {\bibinfo  {journal}
  {arXiv:1502.06263}\ } (\bibinfo {year} {2015})}\BibitemShut {NoStop}%
\bibitem [{\citenamefont {Kalinikos}(1981)}]{Kalinikos1981}%
  \BibitemOpen
  \bibfield  {author} {\bibinfo {author} {\bibfnamefont {B.~A.}\ \bibnamefont
  {Kalinikos}},\ }\href@noop {} {\bibfield  {journal} {\bibinfo  {journal}
  {Sov. Phys. J.}\ }\textbf {\bibinfo {volume} {24}},\ \bibinfo {pages} {718}
  (\bibinfo {year} {1981})}\BibitemShut {NoStop}%
\bibitem [{\citenamefont {Gurevich}\ and\ \citenamefont
  {Melkov}(1996)}]{Gurevich1996}%
  \BibitemOpen
  \bibfield  {author} {\bibinfo {author} {\bibfnamefont {A.}~\bibnamefont
  {Gurevich}}\ and\ \bibinfo {author} {\bibfnamefont {G.}~\bibnamefont
  {Melkov}},\ }\href@noop {} {\emph {\bibinfo {title} {Magnetization
  Oscillations and Waves}}}\ (\bibinfo  {publisher} {CRC Press, Inc},\ \bibinfo
  {year} {1996})\BibitemShut {NoStop}%
\bibitem [{\citenamefont {Morris}\ \emph {et~al.}(2017)\citenamefont {Morris},
  \citenamefont {van Loo}, \citenamefont {Kosen},\ and\ \citenamefont
  {Karenowska}}]{Morris2017}%
  \BibitemOpen
  \bibfield  {author} {\bibinfo {author} {\bibfnamefont {R.~G.~E.}\
  \bibnamefont {Morris}}, \bibinfo {author} {\bibfnamefont {A.~F.}\
  \bibnamefont {van Loo}}, \bibinfo {author} {\bibfnamefont {S.}~\bibnamefont
  {Kosen}}, \ and\ \bibinfo {author} {\bibfnamefont {A.~D.}\ \bibnamefont
  {Karenowska}},\ }\href {https://doi.org/10.1038/s41598-017-11835-4}
  {\bibfield  {journal} {\bibinfo  {journal} {Scientific Reports}\ }\textbf
  {\bibinfo {volume} {7}},\ \bibinfo {pages} {11511} (\bibinfo {year}
  {2017})}\BibitemShut {NoStop}%
\bibitem [{\citenamefont {Stancil}\ and\ \citenamefont
  {Prabhakar}(2009)}]{Stancil2009}%
  \BibitemOpen
  \bibfield  {author} {\bibinfo {author} {\bibfnamefont {D.~D.}\ \bibnamefont
  {Stancil}}\ and\ \bibinfo {author} {\bibfnamefont {A.}~\bibnamefont
  {Prabhakar}},\ }\href {\doibase 10.1007/978-0-387-77865-5} {\emph {\bibinfo
  {title} {Spin Waves}}}\ (\bibinfo  {publisher} {Springer Nature},\ \bibinfo
  {year} {2009})\BibitemShut {NoStop}%
\bibitem [{\citenamefont {O'Keeffe}\ and\ \citenamefont
  {Patterson}(1978)}]{Okeeffe1978}%
  \BibitemOpen
  \bibfield  {author} {\bibinfo {author} {\bibfnamefont {T.~W.}\ \bibnamefont
  {O'Keeffe}}\ and\ \bibinfo {author} {\bibfnamefont {R.~W.}\ \bibnamefont
  {Patterson}},\ }\href {\doibase http://dx.doi.org/10.1063/1.325522}
  {\bibfield  {journal} {\bibinfo  {journal} {J. Appl. Phys.}\ }\textbf
  {\bibinfo {volume} {49}},\ \bibinfo {pages} {4886} (\bibinfo {year}
  {1978})}\BibitemShut {NoStop}%
\bibitem [{\citenamefont {Yukawa}\ \emph {et~al.}(1977)\citenamefont {Yukawa},
  \citenamefont {ichi Yamada}, \citenamefont {Abe},\ and\ \citenamefont {ichi
  Ikenoue}}]{yukawa1977}%
  \BibitemOpen
  \bibfield  {author} {\bibinfo {author} {\bibfnamefont {T.}~\bibnamefont
  {Yukawa}}, \bibinfo {author} {\bibfnamefont {J.}~\bibnamefont {ichi Yamada}},
  \bibinfo {author} {\bibfnamefont {K.}~\bibnamefont {Abe}}, \ and\ \bibinfo
  {author} {\bibfnamefont {J.}~\bibnamefont {ichi Ikenoue}},\ }\href
  {http://stacks.iop.org/1347-4065/16/i=12/a=2187} {\bibfield  {journal}
  {\bibinfo  {journal} {Jpn. J. Appl. Phys.}\ }\textbf {\bibinfo {volume}
  {16}},\ \bibinfo {pages} {2187} (\bibinfo {year} {1977})}\BibitemShut
  {NoStop}%
\bibitem [{\citenamefont {Guslienko}\ \emph {et~al.}(2003)\citenamefont
  {Guslienko}, \citenamefont {Chantrell},\ and\ \citenamefont
  {Slavin}}]{Guslienko2003}%
  \BibitemOpen
  \bibfield  {author} {\bibinfo {author} {\bibfnamefont {K.~Y.}\ \bibnamefont
  {Guslienko}}, \bibinfo {author} {\bibfnamefont {R.~W.}\ \bibnamefont
  {Chantrell}}, \ and\ \bibinfo {author} {\bibfnamefont {A.~N.}\ \bibnamefont
  {Slavin}},\ }\href@noop {} {\bibfield  {journal} {\bibinfo  {journal} {Phys.
  Rev. B}\ }\textbf {\bibinfo {volume} {68}},\ \bibinfo {pages} {024422}
  (\bibinfo {year} {2003})}\BibitemShut {NoStop}%
\bibitem [{\citenamefont {Serga}\ \emph {et~al.}(2010)\citenamefont {Serga},
  \citenamefont {Chumak},\ and\ \citenamefont {Hillebrands}}]{Serga2010}%
  \BibitemOpen
  \bibfield  {author} {\bibinfo {author} {\bibfnamefont {A.~A.}\ \bibnamefont
  {Serga}}, \bibinfo {author} {\bibfnamefont {A.~V.}\ \bibnamefont {Chumak}}, \
  and\ \bibinfo {author} {\bibfnamefont {B.}~\bibnamefont {Hillebrands}},\
  }\href {\doibase 10.1088/0022-3727/43/26/264002} {\bibfield  {journal}
  {\bibinfo  {journal} {Journal of Physics D: Applied Physics}\ }\textbf
  {\bibinfo {volume} {43}},\ \bibinfo {pages} {264002} (\bibinfo {year}
  {2010})}\BibitemShut {NoStop}%
\bibitem [{\citenamefont {Anderson}(1964)}]{Anderson1964}%
  \BibitemOpen
  \bibfield  {author} {\bibinfo {author} {\bibfnamefont {E.~E.}\ \bibnamefont
  {Anderson}},\ }\href {\doibase 10.1103/PhysRev.134.A1581} {\bibfield
  {journal} {\bibinfo  {journal} {Phys. Rev.}\ }\textbf {\bibinfo {volume}
  {134}},\ \bibinfo {pages} {A1581} (\bibinfo {year} {1964})}\BibitemShut
  {NoStop}%
\bibitem [{\citenamefont {Hansen}\ \emph {et~al.}(1974)\citenamefont {Hansen},
  \citenamefont {Roeschmann},\ and\ \citenamefont {Tolksdorf}}]{Hansen1974a}%
  \BibitemOpen
  \bibfield  {author} {\bibinfo {author} {\bibfnamefont {P.}~\bibnamefont
  {Hansen}}, \bibinfo {author} {\bibfnamefont {P.}~\bibnamefont {Roeschmann}},
  \ and\ \bibinfo {author} {\bibfnamefont {W.}~\bibnamefont {Tolksdorf}},\
  }\href {\doibase 10.1063/1.1663657} {\bibfield  {journal} {\bibinfo
  {journal} {J. Appl. Phys.}\ }\textbf {\bibinfo {volume} {45}},\ \bibinfo
  {pages} {2728} (\bibinfo {year} {1974})}\BibitemShut {NoStop}%
\bibitem [{\citenamefont {Schiffer}\ \emph {et~al.}(1995)\citenamefont
  {Schiffer}, \citenamefont {Ramirez}, \citenamefont {Huse}, \citenamefont
  {Gammel}, \citenamefont {Yaron}, \citenamefont {Bishop},\ and\ \citenamefont
  {Valentino}}]{Schiffer1995}%
  \BibitemOpen
  \bibfield  {author} {\bibinfo {author} {\bibfnamefont {P.}~\bibnamefont
  {Schiffer}}, \bibinfo {author} {\bibfnamefont {A.~P.}\ \bibnamefont
  {Ramirez}}, \bibinfo {author} {\bibfnamefont {D.~A.}\ \bibnamefont {Huse}},
  \bibinfo {author} {\bibfnamefont {P.~L.}\ \bibnamefont {Gammel}}, \bibinfo
  {author} {\bibfnamefont {U.}~\bibnamefont {Yaron}}, \bibinfo {author}
  {\bibfnamefont {D.~J.}\ \bibnamefont {Bishop}}, \ and\ \bibinfo {author}
  {\bibfnamefont {A.~J.}\ \bibnamefont {Valentino}},\ }\href {\doibase
  10.1103/physrevlett.74.2379} {\bibfield  {journal} {\bibinfo  {journal}
  {Phys. Rev. Lett.}\ }\textbf {\bibinfo {volume} {74}},\ \bibinfo {pages}
  {2379} (\bibinfo {year} {1995})}\BibitemShut {NoStop}%
\bibitem [{\citenamefont {Schiffer}\ \emph {et~al.}(1994)\citenamefont
  {Schiffer}, \citenamefont {Ramirez}, \citenamefont {Huse},\ and\
  \citenamefont {Valentino}}]{Schiffer1994}%
  \BibitemOpen
  \bibfield  {author} {\bibinfo {author} {\bibfnamefont {P.}~\bibnamefont
  {Schiffer}}, \bibinfo {author} {\bibfnamefont {A.~P.}\ \bibnamefont
  {Ramirez}}, \bibinfo {author} {\bibfnamefont {D.~A.}\ \bibnamefont {Huse}}, \
  and\ \bibinfo {author} {\bibfnamefont {A.~J.}\ \bibnamefont {Valentino}},\
  }\href {\doibase 10.1103/physrevlett.73.2500} {\bibfield  {journal} {\bibinfo
   {journal} {Phys. Rev. Lett.}\ }\textbf {\bibinfo {volume} {73}},\ \bibinfo
  {pages} {2500} (\bibinfo {year} {1994})}\BibitemShut {NoStop}%
\bibitem [{\citenamefont {Deen}\ \emph {et~al.}(2015)\citenamefont {Deen},
  \citenamefont {Florea}, \citenamefont {Lhotel},\ and\ \citenamefont
  {Jacobsen}}]{Deen2015}%
  \BibitemOpen
  \bibfield  {author} {\bibinfo {author} {\bibfnamefont {P.~P.}\ \bibnamefont
  {Deen}}, \bibinfo {author} {\bibfnamefont {O.}~\bibnamefont {Florea}},
  \bibinfo {author} {\bibfnamefont {E.}~\bibnamefont {Lhotel}}, \ and\ \bibinfo
  {author} {\bibfnamefont {H.}~\bibnamefont {Jacobsen}},\ }\href {\doibase
  10.1103/physrevb.91.014419} {\bibfield  {journal} {\bibinfo  {journal} {Phys.
  Rev. B}\ }\textbf {\bibinfo {volume} {91}} (\bibinfo {year} {2015}),\
  10.1103/physrevb.91.014419}\BibitemShut {NoStop}%
\bibitem [{\citenamefont {Paddison}\ \emph {et~al.}(2015)\citenamefont
  {Paddison}, \citenamefont {Jacobsen}, \citenamefont {Petrenko}, \citenamefont
  {Fernandez-Diaz}, \citenamefont {Deen},\ and\ \citenamefont
  {Goodwin}}]{Paddison2015}%
  \BibitemOpen
  \bibfield  {author} {\bibinfo {author} {\bibfnamefont {J.~A.~M.}\
  \bibnamefont {Paddison}}, \bibinfo {author} {\bibfnamefont {H.}~\bibnamefont
  {Jacobsen}}, \bibinfo {author} {\bibfnamefont {O.~A.}\ \bibnamefont
  {Petrenko}}, \bibinfo {author} {\bibfnamefont {M.~T.}\ \bibnamefont
  {Fernandez-Diaz}}, \bibinfo {author} {\bibfnamefont {P.~P.}\ \bibnamefont
  {Deen}}, \ and\ \bibinfo {author} {\bibfnamefont {A.~L.}\ \bibnamefont
  {Goodwin}},\ }\href {\doibase 10.1126/science.aaa5326} {\bibfield  {journal}
  {\bibinfo  {journal} {Science}\ }\textbf {\bibinfo {volume} {350}},\ \bibinfo
  {pages} {179} (\bibinfo {year} {2015})}\BibitemShut {NoStop}%
\end{thebibliography}%

\end{document}